# Complete DFM Model for High-Performance Computing SoCs with Guard Ring and Dummy Fill Effect

Chun-Chen Liu, Oscar Lau, Jason Y. Du
University of California, Los Angeles yuandu@ucla.edu

*Abstract*— **For nanotechnology, the semiconductor device is scaled down dramatically with additional strain engineering for device enhancement, the overall device characteristic is no longer dominated by the device size but also circuit layout. The higher order layout effects, such as well proximity effect (WPE), oxide spacing effect (OSE) and poly spacing effect (PSE), play an important role for the device performance, it is critical to understand Design for Manufacturability (DFM) impacts with various layout topology toward the overall circuit performance. Currently, the layout effects (WPE, OSE and PSE) are validated through digital standard cell and analog differential pair test structure. However, two analog layout structures: the guard ring and dummy fill impact are not well studied yet, then, this paper describes the current mirror test circuit to examine the guard ring and dummy fills DFM impacts using TSMC 28nm HPM process.**

## I. INTRODUCTION

With advance of nanotechnology, the semiconductor device is scaled down rapidly with additional strain engineering for device enhancement, the overall device characteristic is no longer dominated by the device size but also layout effects (WPE, OSE and PSE) [1]. It is critical to understand Design for Manufacturability (DFM) impacts with various layout topology toward the overall circuit performance. Currently, the digital standard cell and analog differential pair layout test structure are implemented to validate the layout effects. However, two important analog circuit layout topology: guard ring and dummy fill impact are not well studied yet. Therefore, this paper describes a current mirror test circuit to examine the guard ring and dummy fills DFM impacts using TSMC 28nm HPM process.

For analog design, the circuit performance is highly dependent on the device matching, then, the centroid topology is often chosen to minimize the layout environmental variation. The transistors are arranged symmetrically in centroid style where all the devices suffer from the same physical and electrical impacts from all directions. The centroid topology focuses the active devices layout impacts only, the guard ring protection and dummy fill layout structures are not fully taken into consideration yet. At a result, the modified current mirror configurations are implemented to explore various guard ring and dummy fill DFM impacts.

## II. MODIFIED TEST STRUCTURE

In Figure 1, it is shown the conventional current mirror [2] centroid layout topology, the multi-finger devices: MA and MB are placed alternatively and symmetrically, the individual transistor experience same layout impacts, they suffer same physical and electrical impacts from all direction [3]. This layout topology is originally developed to minimize the angular implant doping variation, it is further enforced to reduce layout effects (WPE, OSE and PSE) impacts since 45nm process [4][5]. This paper focuses on the fine centroid layout style rather than the coarse one where a group of transistors are arranged symmetrically to minimize high interconnect RC parasitic impacts.

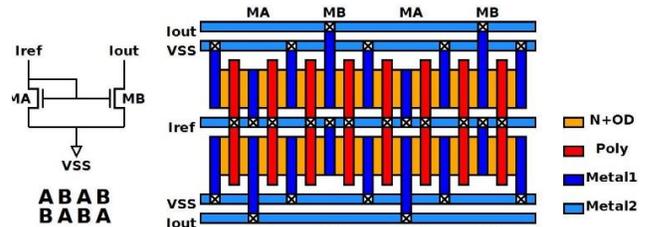

Figure 1. NMOS Current Mirror Fine Centroid Topology

In order to isolate the current mirror from other impacts, the active devices are protected by guard ring and diffusion (OD)/poly (PO) dummy. The guard ring is typically used to protect the active devices from latch-up and noise interference where P+ guard ring with VSS connection protects NMOS active devices, PMOS active devices surrounds with N+ guard ring connected to VDD as shown in Figure 2. There are two kinds of guard ring: single and double guard ring where the single guard ring employs either P+ or N+ guard ring only, the double guard ring mixes with both P+ and N+ guard ring.

Currently, the simulation model only considers the active device layout effect impacts, it ignores the physical and electrical impacts introduced by guard ring. Only diffusion spacing between active devices and guard ring are considered in simulation using OSE model. The guard ring diffusion width and P+/N+ implant type both contribute to the device mobility changes. Therefore, we propose to modify the original OSE model [6] by introducing effective STI width (STIWeff) parameter. As a first-order model, we set a threshold of OD width of single guard ring (ODWth) when guard ring effect comes into play. The value of ODWth can be found by experiment. If ODW is smaller than ODWth, then the STIeff is defined as

$$STIW_{eff} = STIW \times \left(1 + K \frac{ODW_{th}}{ODW}\right) \quad (1)$$

where K is curve fitting parameter.

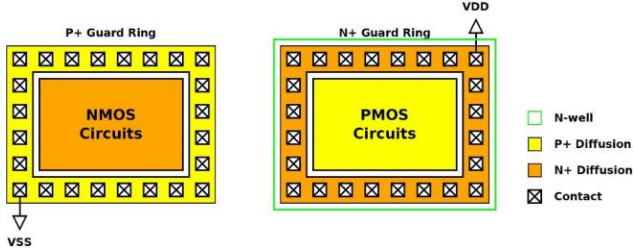

Figure 2. P+/N+ Guard Ring

Dummy fill is typically related with three different type dummy: diffusion, poly and metal. This paper mainly focuses on diffusion (OD) dummy fill because the diffusion uniformity is directly related with Shallow Trench Isolation (STI) and Rapid Temperature Annealing (RTA) process that linked with the transistors formation. The poly (PO) dummy fill is critical for 3D FinFet technology because the height of FinFet is directly dependent on the PO dummy density. Finally, the metal dummy is closely linked to Chemical Mechanical Polishing (CMP) process and directly impacts toward interconnect RC parasitic. For current test chip implementation, it is divided into two level dummy fill. The first level dummy structure is similar to current mirror with same dimension and diffusion type, the second level dummy structure is used to examine different OD fill impacts as shown in Figure 3.

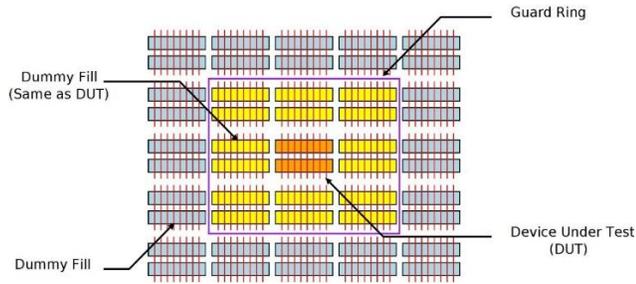

Figure 3. Device under Test (DUT) Structure

Most of OD dummy fill is limited to N+OD dummy fill in order to simplify the chip implementation, it further explores with P+OD and mixed N+OD/P+OD dummy fill impacts toward different current mirror configurations. The conventional fine centroid PMOS/NMOS current mirrors are chosen with N+OD dummy fills as reference circuit for comparison because N+OD dummy fill is commonly offered by foundry. It is easy to implement compared with P+OD and mixed one. For test chip implementation, it focuses on the different type of diffusion impacts toward current mirror performance. All the poly, diffusion and metal density are same for all test structures, except the type of diffusion. Three type diffusion: N+OD, P+OD are mixed one are implemented. The type of diffusion is highly related with Rapid Temperature Annealing (RTA) process for device threshold voltage and leakage variation [7]. The effect threshold voltage can be expressed as:

$$V_{teff} = V_{tref}\big(1 + f(D_{NOD}, D_{POD})\big) \quad (2)$$

where DNOD and DPOD are dummy density for N+OD and P+OD within 100 μm x 100 μm window around MOSFETs, respectively, and f(DNOD, DPOD) is a look-up table function which is obtained by measurement. Therefore, the test structure is used to study threshold voltage changes with different type of dummy diffusion fills.

### III. TEST CHIP RESULTS

In order to validate analog test structure, various current mirrors are implemented using TSMC 28nm HPM process as shown in Figure 4. It includes five PMOS (Table 1) and five NMOS (Table 2) current mirrors. The distance between guard ring and active devices is 1μm. For single guard ring, the OD widths in guard ring are 0.14 μm (1X) and 0.28 μm (2X), respectively. For double guard ring, the OD width is 0.28 μm for both P-type and N-type guard ring. The distance between N-type and P-type guard ring is 0.4 μm. For OD dummy fill, the total OD density is over 50% within 100 μm x 100 μm window. For N+/P+ OD dummy fill, half of OD dummy is P-type and the other half is N-type.

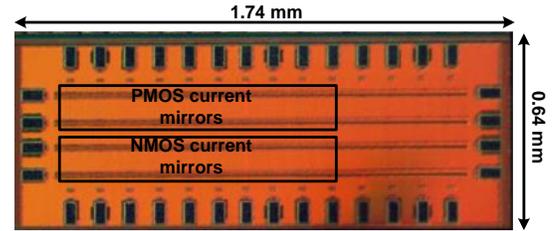

Figure 4. Guard Ring and Dummy Fill Die photo

| Type | Guard Ring | Dummy | Simulated Ratio | Measured Ratio |
|---|---|---|---|---|
| PMOS | Double | P+OD | 1.00 | 1.00 |
| PMOS | Single 1X | P+OD | 1.00 | 1.01 |
| PMOS | Single 2X | P+OD | 1.00 | 1.05 |
| PMOS | Double | N+OD | 1.00 | 0.99 |
| PMOS | Double | N+/P+OD | 1.00 | 0.99 |

Table 1. PMOS Current Mirror Measurement

From Table 1, the single guard ring has 1%~5% larger current compared with double guard ring because the PMOS guard ring introduces additional tensile stress to enhance mobility, and the tensile stress is further increased with wider guard ring (i.e. 2X). This impact is predicted by proposed modified OSE model. The comparison of simulation with original OSE model, with proposed OSE model and measurement is shown in Fig. 4. For double guard ring, the P+/N+ guard rings offset each other performance enhancement. Regard to different OD fill, the impacts are relatively minor with 1% performance difference. This means

f(D_NOD,D_POD ) has small value for PMOS.

| Type | Guard Ring | Dummy | Simulated Ratio | Measured Ratio |
|---|---|---|---|---|
| NMOS | Double | N+OD | 1.00 | 1.00 |
| NMOS | Single 1X | N+OD | 1.00 | 1.01 |
| NMOS | Single 2X | N+OD | 1.00 | 1.05 |
| NMOS | Double | P+OD | 1.00 | 1.09 |
| NMOS | Double | N+P+OD | 1.00 | 1.10 |

Table 2. NMOS Current Mirror Measurement

From Table 2, NMOS current mirror performance is different from PMOS one due to compressive stress that degrades the device driving capability. The single guard ring shows the performance enhancement similar to PMOS one. P+OD and mixed OD dummy fills increase the current by about 10%.

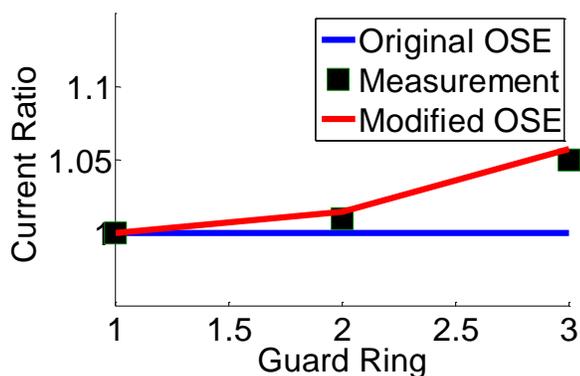

Figure 5. Comparison of original OSE model, measurement and modified OSE model

## IV. CONCLUSION

The guard ring and OD dummy fill effect on performance of MOSFET is studied in this paper. The silicon measurement and simulation result difference is as high as 10% with guard ring and dummy fill, compared the original model with the modified model. The preliminary guard ring and OD dummy fill models are proposed to improve the simulation accuracy. For future development, more test keys are requirement to estimate the OD density gradient impacts and how to apply guard ring to protect the active circuit from the neighbor passive resistor and capacitor.